\documentclass[aps, prb, twocolumn, superscriptaddress]{revtex4-2}
\usepackage[]{physics}
\usepackage[]{amsmath}
\usepackage{amssymb}
\usepackage{mathrsfs}

\usepackage[T1]{fontenc}
\usepackage{fontawesome}

\usepackage[]{graphicx}
\graphicspath{{src/figures/}}
\usepackage[]{mathtools}
\usepackage[]{bm}
\usepackage[]{braket}
\usepackage{esint}
\usepackage{ulem}
\usepackage{cancel}
\usepackage[colorlinks=true]{hyperref}
\usepackage[caption=false]{subfig}

\newcommand{\figref}[1]{\figurename~\ref{#1}}

\usepackage{cases}
\usepackage[]{comment}

\begin{document}

\title{
  Noncontact frictional force between surfaces by peristaltic permittivity modulation 
}

\author{Daigo Oue}
\email{daigo.oue@gmail.com}
\affiliation{Instituto Superior T\'{e}cnico, University of Lisbon, 1049-001 Lisboa, Portugal}
\affiliation{The Blackett Laboratory, Imperial College London, London SW7 2AZ, United Kingdom}
\author{Kun Ding}
\affiliation{Department of Physics, State Key Laboratory of Surface Physics, and Key Laboratory of Micro and Nano Photonic Structures (Ministry of Education), Fudan University, Shanghai 200438, China}
\author{J. B. Pendry}
\affiliation{The Blackett Laboratory, Imperial College London, London SW7 2AZ, United Kingdom}

\date{\today}

\begin{abstract}
  In this study, we reveal noncontact frictional forces between surfaces in the presence of peristaltic permittivity modulation.
  Our setup comprises a conducting medium, an air gap, and a dielectric substrate on which we have a space-time-modulated grating that emits electromagnetic radiation.
  The radiation receives energy and momentum from the grating,
  which is eventually absorbed by the conducting medium or propagates away from the grating on the dielectric side,
  resulting in electromagnetic power loss and lateral forces at the surfaces.
\end{abstract}

\maketitle

\section{Introduction}
Periodic structures have been widely used to confine electromagnetic fields~\cite{yablonovitch1994photonic,knight2003photonic,soljavcic2004enhancement,sakoda2004optical}.
The structure modulates the electromagnetic density of states, as the crystal structure in solid-states materials does the electron density of states, and the field can be spatially confined;
thereby, strong field-matter interaction is induced~\cite{maldovan2006simultaneous,eichenfield2009optomechanical,muallem2016strong,zhang2018photonic}. 
That is why such a system has been applied to biological sensors~\cite{skivesen2007photonic,inan2017photonic} and optical devices such as lasers~\cite{painter1999room,lonvcar2002low,altug2006ultrafast} and detectors~\cite{rosenberg2010design,shiue2013enhanced,schuler2018graphene}.

Due to the advances in finer structure fabrication, it has been more popular to design structures rather than chemical compositions to control electromagnetic waves, where the structured system can be treated as an effective medium~\cite{garnett1904xii,garnett1906vii,bruggeman1935berechnung,bruggeman1936berechnung} and is called metamaterials.
Appropriately designing the structure enables negative refraction~\cite{smith2000composite,smith2000negative,shelby2001experimental,shelby2001microwave,smith2004metamaterials}, perfect lens~\cite{pendry2000negative,pendry2002near,smith2004metamaterials,melville2005super,fang2005sub}, and wave cloaking~\cite{ochiai2008novel,xu2013transformation,chen2010transformation,chen2011conformal,xu2015conformal}.

Although the scalable fabrication of three-dimensional fine structure is still under the development and a hard task, particularly in the optical frequency range~\cite{takeyasu2016invited,taguchi2018extraction}, the two-dimensional version of metamaterials, also known as metasurfaces, has been within the reach of experiments even at the optical frequencies.
The metasurfaces have been applied, e.g., to harnessing optical angular momenta~\cite{li2013spin,chen2015creating,devlin2017spin,li2017nonlinear,sroor2020high,zhang2022chiral} and to realising ultrathin lens, which can focus light waves without suffering from various types of abberations~\cite{aieta2012aberration,yu2014flat,wen2018geometric}.

Recently, investigating time-varying media in addition to the structured media, which involves earlier studies~\cite{simon1960action,oliner1961wave,cassedy1963dispersion,cassedy1967dispersion,winn1999interband,biancalana2007dynamics}, has drawn more and more interest.
This is partly because of 
their nontrivial topology~\cite{fang2012photonic,fang2012realizing,lustig2018topological,wang2020floquet,sounas2017non},
application to nonreciprocal light propagation~\cite{sounas2017non,shaltout2015time,mazor2019nonreciprocal}, 
light compression and amplification~\cite{galiffi2019broadband,pendry2021gainJOSAB,pendry2022photon,pendry2022crossing},
the emission of radiation \cite{belgiorno2010hawking,sloan2021casimir,sloan2022controlling,oue2022cerenkov},
and the enhancements of free-electron and dipole radiations~\cite{hu2021free,dikopoltsev2022light,lyubarov2022amplified}.

Most of these papers focused on the propagation and confinement of electromagnetic fields in time-varying, structures media.
In this work, we address the energy consumption and electromagnetic forces associated with the radiation emission in such a medium.

\begin{figure}[tbp]
  \centering
  \includegraphics[width=\linewidth]{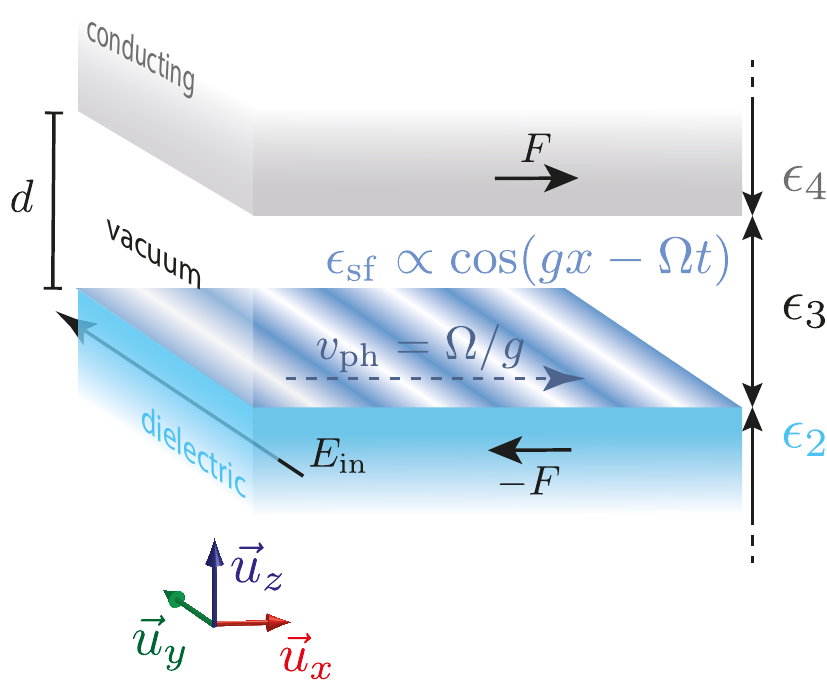}
  \caption{
    The schematic image of the setup analysed in this study.
    We will consider electromagnetic forces $F$ exerted between two surfaces separated by a vacuum gap ($\epsilon _ 3 = 1$) with width $d$.
    On the lower surface, we have a space-time modulated grating $\epsilon _ \mathrm{sf}$ performing peristaltic motion with a peristaltic speed $v _ \mathrm{ph} := \Omega/g$.
    The upper medium is lossy, characterised by the permittivity of conducting material \eqref{eq:eps4}.
    We assume that the lower medium is dielectric,
    $\epsilon _ 2 = \mathrm{const.}$,
    and $\mu = 1$ everywhere for simplicity.
    A DC voltage $E _ \mathrm{in}$ is applied to the lower medium.
    The temporal modulation operates in the GHz frequency range.
  }
  \label{fig:setup}
\end{figure}

Our setup is composed of a lossy medium, a vacuum gap, and a dielectric substrate on which a space-time modulated grating is placed (see \figref{fig:setup}).
The grating is modelled by the effective surface permittivity,
\begin{align}
  \epsilon _ \mathrm{sf} =
  \epsilon_1(1 + 2\alpha \cos \mathbf{q} \cdot \mathbf{x}) =
  \epsilon_1[1 + 2\alpha \cos (gx - \Omega t)],
  \label{eq:eps_sf}
\end{align}
where $\epsilon _ 1$ is the effective surface permittivity in the absence of modulation, $\alpha$ is the strength of the modulation, and the spatial and temporal modulation frequencies is denoted by $g$ and $\Omega$.
In particular, we consider temporal modulation operating in the GHz frequency range below.
Note that we have introduced three-component quantities $\mathbf{q} := (g, 0, i\Omega/c)$ and $\mathbf{x} := (x, y, ict)$,
and $c \equiv 1/\sqrt{\mu _ 0 \epsilon _ 0}$ is the speed of light.
The grating performs peristaltic motion with a peristaltic velocity $v _ \mathrm{ph} := \Omega/g$.
The lossy medium filling the upper region, $o _ 4 = \qty{z|\ d < z}$, is modeled by the permittivity of conducting material,
\begin{align}
  \epsilon_{4} = 1+i\frac{\kappa/\epsilon_0}{\omega},
  \label{eq:eps4}
\end{align}
where $\kappa$ is the conductivity.
Note that, more generally, conducting materials can be modeled by the Drude permittivity, $1 - \omega _ \mathrm{p} ^ 2/(\omega ^ 2 + i\omega\Gamma)$, where $\omega _ \mathrm{p}$ and $\Gamma$ are the plasma frequency and the damping constant;
however, we focus on the damping is significant $\omega \ll \Gamma$, where the Drude permittivity is reduced to Eq.~\eqref{eq:eps4}, to efficiently capture the radiation emitted from the grating and the associated forces as we will see below.
We have constant permittivities $\epsilon _ 3$ in the gap region, $o _ 3 = \qty{z|\ 0 < z < d}$, and $\epsilon _ 2$ in the lower region, $o _ 2 = \qty{z|\ z < 0}$.

Our input electrostatic field is uniformly applied to the lower dielectric medium.
The space-time grating provides not only momentum $g$ but also energy $\Omega$ to the electromagnetic system.
While conventional gratings, which supply momentum, only redistribute electrostatic fields over space,
the space-time grating emits electromagnetic radiation in its lower and upper sides if the field receives sufficient energy.
If the radiation emitted by the grating is absorbed by the conducting medium or propagates away from the grating,
there will be power loss and associated forces at the surfaces,
which we are going to calculate in this work.

Our consideration is closely related to noncontact friction problem.
In general, noncontact friction can be viewed as an interaction force due to their correlation mediated by a field between two relatively moving bodies.
There are various examples: frictional forces mediated by (i) static electromagnetic fields~\cite{persson1998theory,stipe2001noncontact,chumak2004effects,kuehn2006dielectric,saitoh2010gigantic, she2012noncontact, den2015spin, de2017dissipation, heritier2021spatial} and (i\hspace{-.05em}i) dynamical electromagnetic fields~\cite{pendry1997shearing,pendry1998can,philbin2009no,pendry2010quantum,leonhardt2010comment,pendry2010reply,silveirinha2014theory,milton2016reality,dedkov2017fluctuation,reiche2022wading}.
In the second example, the mediator field could be thermal or quantum ones;
thereby, the frictional force stems from the quantum uncertainty and remains even at the zero-temperature and also called quantum friction.
Our setup is similar to the first example owing to the fact that our grating is electrically polarised, assisted by the DC voltage, and an electrostatic field clings to it.
On the other hand, they are different in that our electric polarisation is conveyed by the peristaltic modulation instead of physically moving the medium.
As this motion is non-physical, the resultant velocity can surpass that of light, and the mediator field surrounding the grating can be of the \v{C}erenkov variety, which is dynamical rather than static.
This makes our setup similar to the second one.

This paper is organized as follows.
In Sec.~I\hspace{-.1em}I, we summarise how to calculate the scattering of an electromagnetic field at the space-time modulated grating and multiple reflections between the grating and the second surface. 
In Sec.~I\hspace{-.1em}I\hspace{-.1em}I, we numerically evaluate the power of emission from the grating.
The electromagnetic force associated with the radiation emission from the grating is analysed in Sec.~I\hspace{-.1em}V.
The discussion and the conclusion are drawn in Sec.~V.

\section{Multiple reflections between the surfaces}
To calculate the power loss and frictional forces,
we first compute the field amplitude with the multiple reflection formalism,
where we need the reflection and transmission matrices of each surface.
We can safely use the conventional Fresnel coefficients for the upper flat surface.
The reflection and transmission coefficients of the lower surface can be obtained from the boundary conditions in the presence of a grating-induced source~\cite{oue2021calculating,oue2022cerenkov} as we will summarise below.

The electric and magnetic fields evaluated at a position $(\vb{x},z)$ can be expanded in plane waves,
\begin{align}
  \vec{\mathcal{E}} (\vb{x},z) &=
  \sum _ {\sigma, \tau, m}
  E _ {\vb{k} _ m} ^ {(\tau) \sigma} \vec{\varepsilon} _ {\vb{k} _ m} ^ {(\tau) \sigma}(z)
  e ^ {i \vb{k} _ m \cdot \vb{x}},
  \label{eq:E}
  \\
  \vec{\mathcal{H}} (\vb{x},z) &=
  \sum _ {\sigma, \tau, m}
  H _ {\vb{k} _ m} ^ {(\tau) \sigma} \vec{h} _ {\vb{k} _ m} ^ {(\tau) \sigma}(z)
  e ^ {i \vb{k} _ m \cdot \vb{x}},
  \label{eq:H}
\end{align}
where we substitute $\vb{k} _ m = m\vb{q}$,
one of the superscripts, $\sigma \in \qty{+,-}$, characterises the propagation direction,
and the other, $\tau \in \qty{2,3,4}$, specifies the medium.
The electric and magnetic field amplitudes are associated with each other via the impedance,
\begin{align}
  \frac{E _ {\vb{k}} ^ {(\tau) \sigma}}{H _ {\vb{k}} ^ {(\tau) \sigma}} =
  \frac{Z _ 0}{\sqrt{\epsilon _ \tau}} \frac{|\omega/c|}{\sqrt{|K _ {\vb{k}} ^ {(\tau)}| ^ 2 + k _ \parallel ^ 2}} =:
  Z _ {\vb{k}} ^ {(\tau)},
\end{align}
where we have introduced the free space impedance,
$Z _ 0 \equiv \sqrt{\mu _ 0/\epsilon _ 0}$,
the wave number parallel to the surfaces,
and the one in the $z$ direction,
\begin{align}
k _ \parallel := \sqrt{k _ x ^ 2 + k _ y ^ 2},
\quad
K _ {\vb{k}} ^ {(\tau)} := \sqrt{\omega ^ 2 \epsilon _ \tau / c ^ 2 - k _ \parallel ^ 2}.
\end{align}
The corresponding wave vector is 
$\vec{k} = \operatorname{sgn}(\omega)
(k _ x \vec{u} _ x + k _ y \vec{u} _ y +
\sigma K _ {\vb{k}} ^ {(\tau)} \vec{u} _ z)$.
The polarisation vectors are defined by
\begin{align}
  \vec{\varepsilon} _ {\vb{k}} ^ {(\tau) \sigma}(z) &:=
  \frac{1}{\sqrt{V}}
  \frac{\vec{k} \times \vec{u} _ z}{|\vec{k} \times \vec{u} _ z|}
  \chi _ \tau (z)
  e ^ {i\sigma K _ {\vb{k}} ^ {(\tau)} z},
  \\
  \vec{h} _ {\vb{k}} ^ {(\tau) \sigma}(z) &:= 
  \frac{1}{\sqrt{V}}
  \frac{\vec{k} \times \vec{k} \times \vec{u} _ z}{|\vec{k} \times \vec{k} \times \vec{u} _ z|}
  \chi _ \tau (z) e ^ {i\sigma K _ {\vb{k}} ^ {(\tau)} z},
\end{align}
where $\vec{u} _ j$ is the unit vector in the $j$-direction ($j = x,y,z$),
and $1/\sqrt{V}$ is the normalisation factor.
We have introduced a characteristic function $\chi _ \tau (z)$ returning 1 if $z \in o _ \tau$, and 0 otherwise;
hence, 
$\chi _ \tau \chi _ {\tau'} \propto \delta _ {\tau \tau'}$.
Note that we shall omit indices and arguments as we may without any possibility of confusion.

The transverse component of the electric field is continuous at the grating plane ($z = 0$),
\begin{align}
  \lim _ {\delta h \downarrow 0} \vec{u} _ y \cdot [\vec{\mathcal{E}}(\mathbf{x}, +\delta h) - \vec{\mathcal{E}}(\mathbf{x},-\delta h)] = 0,
  \label{eq:E_continuity}
\end{align}
while the magnetic field is discontinuous because of the contribution of the grating,
\begin{align}
  \lim _ {\delta h \downarrow 0} \vec{u} _ x \cdot [\vec{\mathcal{H}}(\mathbf{x}, +\delta h) - \vec{\mathcal{H}}(\mathbf{x},-\delta h)] =
  \vec{u} _ y \cdot \frac{\dot{\epsilon} _ \mathrm{sf} \vec{\mathcal{E}} (\mathbf{x},0)}{cZ _ 0}.
  \label{eq:H_continuity}
\end{align}
Applying the Fourier transformation,
these equations (\ref{eq:E_continuity}, \ref{eq:H_continuity}) yield simultaneous equations,
which are inverted to give the reflection and transmission matrices (see Appendix for the detail),
\begin{align}
  \begin{pmatrix}
  \mathbb{E} _ {\vb{k}} ^ {(3)+}\\
  \mathbb{E} _ {\vb{k}} ^ {(2)-}
  \end{pmatrix}
  = 
  \begin{pmatrix}
    \mathsf{R} _ {\vb{k}} & \mathsf{T} _ {\vb{k}} \\
    \mathsf{T} _ {\vb{k}}' & \mathsf{R} _ {\vb{k}}' 
  \end{pmatrix}
  \begin{pmatrix}
  \mathbb{E} _ {\vb{k}} ^ {(3)-}\\
  \mathbb{E} _ {\vb{k}} ^ {(2)+}
  \end{pmatrix}
  \label{eq:scattering_matrix}
\end{align}
where $\mathbb{E} _ {\vb{k}} ^ {(3)-}$ ($\mathbb{E} _ {\vb{k}} ^ {(2)+}$) is the incoming field in the upper (lower) side of the grating.
while $\mathbb{E} _ {\vb{k}} ^ {(3)+}$ ($\mathbb{E} _ {\vb{k}} ^ {(2)-}$) is the outgoing one in each side.
Note that we have collected the field amplitudes in a single column,
$\mathbb{E} _ {\vb{k}} ^ {(\tau) \sigma} := 
\begin{pmatrix}
  \cdots &
  E_{\mathbf{k}_{-1}}^{(\tau) \sigma} &
  E_{\mathbf{k}_{0}}^{(\tau) \sigma} &
  E_{\mathbf{k}_{+1}}^{(\tau) \sigma} &
  \cdots
\end{pmatrix}^\intercal.$
For our consideration, $\mathbb{E} _ {\vb{k}} ^ {(2)+}$ corresponds to the input electrostatic field.

In the absence of the upper surface, the upward (downward) emission from the grating is given by the product of the transmission (reflection) matrix and the input field,
\begin{align}
  \mathbb{E} _ {\vb{k}} ^ {(3)+} =
  \mathsf{T} _ {\vb{k}} \mathbb{E} _ {\vb{k}} ^ {(2)+},
  \quad
  \mathbb{E} _ {\vb{k}} ^ {(2)-} =
  \mathsf{R} _ {\vb{k}} \mathbb{E} _ {\vb{k}} ^ {(2)+},
\end{align}
Inserting the upper surface gets the upward emission back to the grating that reflects the emission again,
and we have countably many reflections between the upper surface and the grating plane.

Making use of the multiple reflection arguments (see Appendix for the detail),
we can expand the field amplitude in powers of the reflection matrices.
We can write the resultant modal amplitudes between the surfaces,
\begin{align}
  \mathbb{E} _ {\vb{k}} ^ {(3)+} =
  \mathsf{S} _ {\vb{k}} \mathsf{T} _ {\vb{k}} \mathbb{E} _ {\vb{k}} ^ {(2)+},
  \quad
  \mathbb{E} _ {\vb{k}} ^ {(3)-} =
  \mathsf{r} _ {\vb{k}} \mathsf{S} _ {\vb{k}} \mathsf{T} _ {\vb{k}} \mathbb{E} _ {\vb{k}} ^ {(2)+},
  \label{eq:E_multiplied}
\end{align}
and the amplitudes in the upper and lower semi-infinite regions,
\begin{align}
  \mathbb{E} _ {\vb{k}} ^ {(4)+} &=
  \mathsf{t} _ {\vb{k}} \mathsf{S} _ {\vb{k}} \mathsf{T} _ {\vb{k}} 
  \mathbb{E} _ {\vb{k}} ^ {(2)+},
  \label{eq:E_multiplied_outer4}
  \\
  \mathbb{E} _ {\vb{k}} ^ {(2)-} &=
  (\mathsf{R} _ {\vb{k}} ' + \mathsf{T} _ {\vb{k}}' \mathsf{r} _ {\vb{k}} \mathsf{S} _ {\vb{k}} \mathsf{T} _ {\vb{k}})
  \mathbb{E} _ {\vb{k}} ^ {(2)+},
  \label{eq:E_multiplied_outer3}
\end{align}
where we have introduced the multiple reflection factor
$\mathsf{S} _ {\vb{k}} := (1-\mathsf{R} _ {\vb{k}} \mathsf{r} _ {\vb{k}})^{-1}$
with the reflection $\mathsf{r} _ {\vb{k}}$ and transmission matrices $\mathsf{t} _ \mathbf{k}$ of the upper flat surface.

\begin{figure}[htbp]
  \centering
  \includegraphics[width=.9\linewidth]{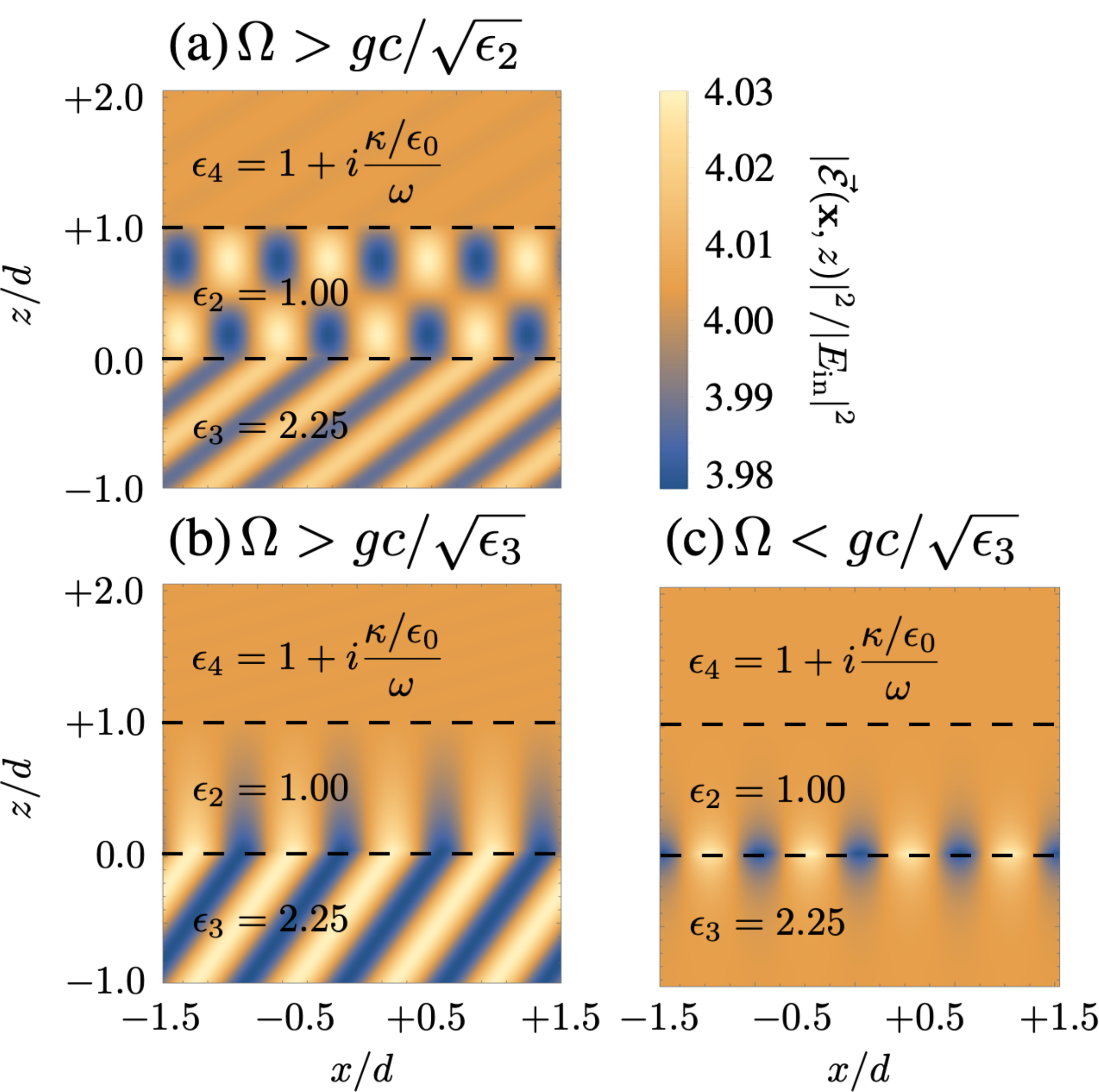}
  \caption{
    Snapshots of the electric field amplitude emitted by the grating at $t=0$.
    The field pattern shifts to the right at the peristaltic velocity $v _ \mathrm{ph}$ along with the grating as time passes.
    The horizontal dashed lines in the figures correspond to the grating plane ($z=0$) and the second surface ($z = d$).
    (a) superluminal regime in both vacuum and dielectric regions ($gc/\sqrt{\epsilon_2} < \Omega$);
    (b) superluminal/superluminal in the vacuum/dielectric region ($gc/\sqrt{\epsilon_3} < \Omega < gc/\sqrt{\epsilon_2}$);
    (c) subluminal in both regions ($\Omega < gc/\sqrt{\epsilon_3}$).
    We have used the following parameters to generate those figures:
    $d = 2.0\ \mathrm{mm}$;
    $\epsilon _ 2 = 1.00$;
    $\epsilon _ 3 = 2.25$;
    $\kappa = 2.0\times10^3\ \mathrm{S/m}$;
    $g = 0.5\ \mathrm{mm ^ {-1}}$;
    $\alpha = -0.19$;
    $\epsilon _ 1 = 3.3 \times 10^{-6}$.
  }
  \label{fig:field_pattern}
\end{figure}

Once the modal amplitudes have been evaluated according to Eqs.~(\ref{eq:E_multiplied}--\ref{eq:E_multiplied_outer3}),
we can reconstruct the field distribution in the real space using the expression~\eqref{eq:E}.
The field patterns generated by various modulations are shown in \figref{fig:field_pattern}.
If the field receives enough energy ($\Omega > gc/\sqrt{\epsilon _ 3}$), it can escape the grating \cite{oue2022cerenkov}.
Since the conducting medium screens the electromagnetic fields,
the upward emission from the grating is reflected back from the upper surface,
and there are standing waves as confirmed in \figref{fig:field_pattern}(a).
As the temporal modulation frequency decreases,
the field can no longer acquire sufficient energy to leave the grating.
In the intermediate regime ($gc/\sqrt{\epsilon _ 3} < \Omega < gc/\sqrt{\epsilon _ 2}$),
the field is confined in the upper side of the grating while propagates away in the lower side [\figref{fig:field_pattern}(b)].
At far lower frequencies ($\Omega < gc/\sqrt{\epsilon _ 3}$),
there is confinement on the both sides [\figref{fig:field_pattern}(c)].

\section{Power loss at the modulated surface}
In the preceding section, we have found the field emission around the grating for various grating parameters.
In this section, we are going to evaluate the field emission power from the grating.
Our starting point is the energy conservation equation.
Taking the time derivative of the electromagnetic energy density,
$U := \qty(
\vec{\mathcal{E}}\cdot \epsilon_0 \epsilon \vec{\mathcal{E}} + 
\vec{\mathcal{H}}\cdot \mu_0 \vec{\mathcal{H}}
)/2$,
we can obtain the energy conservation equation,
\begin{align}
  \pdv{U}{t} + P = \Delta Q,
  \label{eq:conservation}
\end{align}
where we have defined the power loss,
$P := \vec{\mathcal{E}}\cdot \epsilon_0 \dot{\epsilon} \vec{\mathcal{E}}/2,$
and the incoming flux density, 
$\Delta Q := -\nabla \cdot \vec{\mathcal{E}} \times \vec{\mathcal{H}}.$
After cycle-averaging,
$\expval{\ldots} := \int _ 0 ^ {2\pi} \ldots \dd{(\Omega t)}$,
we obtain $\expval{P} = \expval{\Delta Q}$.
Integrating over an infinitesimally thin volume containing the grating
(i.e.~$|x| < \infty$, $|y| < \infty$, and $|z| < \delta h$),
where the permittivity varies in space and time,
we can write
\begin{align}
  \expval{P}
  &= -\lim _ {\delta h \downarrow 0} \qty[
  \iint \limits _ {-\infty} ^ {+\infty} 
  \expval{\vec{\mathcal{E}} \times \vec{\mathcal{H}}} \cdot \vec{u} _ z \dd{x}\dd{y}
  ] _ {z = -\delta h} ^ {z = +\delta h}.
\end{align}
Hence, the average incoming flux is equal to the average power loss.
Substituting Eqs.~(\ref{eq:E}, \ref{eq:H}),
\begin{align}
  \expval{P} = \sum _ {\tau = 2,3} 
  \begin{pmatrix}
    \mathbb{E} _ {\vb{k}} ^ {(\tau)+}
    \\
    \mathbb{E} _ {\vb{k}} ^ {(\tau)-}
  \end{pmatrix} ^ \dagger
  \begin{pmatrix}
    \mathsf{P} _ {\vb{k}\tau} ^ {++} & \mathsf{P} _ {\vb{k}\tau} ^ {+-} 
    \\
    \mathsf{P} _ {\vb{k}\tau} ^ {-+} & \mathsf{P} _ {\vb{k}\tau} ^ {--} 
  \end{pmatrix}
  \begin{pmatrix}
    \mathbb{E} _ {\vb{k}} ^ {(\tau)+}
    \\
    \mathbb{E} _ {\vb{k}} ^ {(\tau)-}
  \end{pmatrix}
  \label{eq:P}
\end{align}
where $\tau = 2\ (3)$ corresponds to the incoming flux from the upper (lower) side,
and we have defined
\begin{align}
  \qty[\mathsf{P} _ {\vb{k} \tau} ^ {\sigma \sigma'}] _ {n n'} =
  \frac{1}{Z _ {\vb{k} _ n} ^ {(\tau)}}
  \frac{(\sigma K _ {\vb{k} _ {n}} ^ {(\tau)} + \sigma' K _ {\vb{k} _ {n}} ^ {(\tau)})/2}{n\Omega/c} 
  \delta _ {n n'}.
  \label{eq:P _ nn'}
\end{align}
We can recognise the numerator of the second fraction in Eq.~\eqref{eq:P _ nn'} returns the real (imaginary) part of the wavenumber in the $z$ direction if $\sigma = \sigma'$ ($\sigma \neq \sigma'$).
The second fraction overall represents the propagation angle (or the field confinement).
Thus, the diagonal (off-diagonal) part in the quadratic form \eqref{eq:P} is proportional to the real (imaginary) part of the wavenumber in the $z$ direction and hence corresponds to the contributions from propagating (evanescent) waves.
The inverse of the impedance multiplied by the field amplitude squared,
$|E _ {\vb{k} _ n} ^ {(\tau)\sigma}| ^ 2/Z _ {\vb{k} _ n} ^ {(\tau)}$,
gives the radiation intensity of the $n$th mode.
From these observations, the quadratic form \eqref{eq:P} is a decomposition of the emission power into the modal intensities with the correction due to the finite propagation angle (or the field confinement).

\begin{figure}[tbp]
  \centering
  \includegraphics[width=\linewidth]{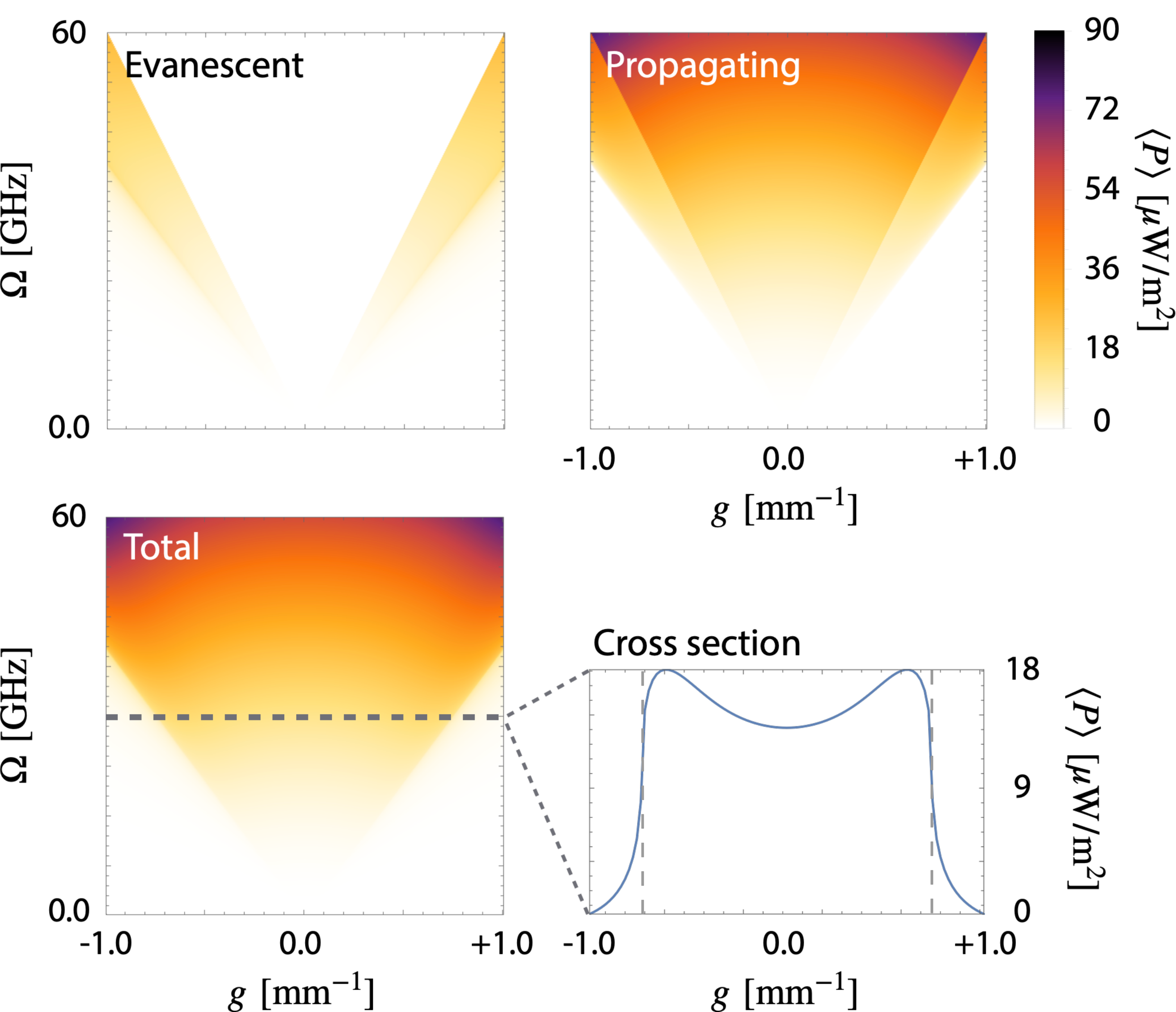}
  \caption{
    Electromagnetic power loss per unit area at the grating.
    The contributions from evanescent (top left), the ones from propagating waves (top right), and the total contribution (bottom left) are shown.
    The bottom right figure is a cross-section corresponding to the grey dashed line ($\Omega = 30\ \mathrm{GHz}$) in the bottom left colour map.
    We have used the following parameters to produce these figures:
    $d = 1.0\ \mathrm{nm}$; 
    $\kappa = 11\ \mathrm{S/m}$;
    $\alpha = -0.19$; 
    $\epsilon_1 = 3.3 \times 10^{-6}$;
    $E _ \mathrm{in} = 0.03\ \mathrm{V/mm}$.
  }
  \label{fig:loss}
\end{figure}
As we have already evaluated the field amplitudes $\mathbb{E} _ {\vb{k}} ^{(\tau)\sigma}$ in the previous section (\ref{eq:E_multiplied}--\ref{eq:E_multiplied_outer3}),
we can compute the power loss with the quadratic form \eqref{eq:P}.
In \figref{fig:loss}, the power loss is plotted as a function of the grating parameters, $g$ and $\Omega$.
The contributions from evanescent and propagating channels and the total contribution are shown.
It is clear that the power loss goes off far from the light cone ($|g| > \Omega\sqrt{\epsilon _ 2}/c$),
where the field acquires much momentum and cannot escape from the grating in either upper and lower media.
We can also recognise sharp changes at the luminal condition (e.g.~vertical dashed lines in the cross-section plot in \figref{fig:loss}) because of the following reasons:
first, the propagating channel opens, and the emitted energy will be taken away in the lower medium;
second, the emitted radiation propagating along the grating strongly interacts with it and receives considerable energy.
The second reason is closely related to the light amplification in the luminal metamaterials~\cite{galiffi2019broadband,pendry2021gain,pendry2021gainJOSAB,galiffi2021photon}.

\section{Frictional force on the second surface}
In the previous sections, we have confirmed that the grating emits electromagnetic radiation, which is reflected back and forth between the grating and the upper surface.
Since the radiation has momentum parallel to the surfaces, it can exert radiation forces which are parallel to the surfaces.
In this section, we are going to evaluate the lateral radiation force on the upper surface.

We integrate the Maxwell stress tensor at the upper surface to calculate the cycle-averaged lateral force,
\begin{align}
  \expval{F} = \lim _ {z' \uparrow d} 
  \iint \limits _ {-\infty} ^ {+\infty} 
  \expval{O _ {xz}(z')} \dd{x}\dd{y},
\end{align}
where the stress tensor is $O _ {jj'} =
\epsilon _ 0 [\mathcal{E} _ j \mathcal{E} _ {j'} - (\vec{\mathcal{E}} \cdot \vec{\mathcal{E}}) \delta _ {jj'} /2] +
\mu _ 0 [\mathcal{H} _ j \mathcal{H} _ {j'} - (\vec{\mathcal{H}} \cdot \vec{\mathcal{H}}) \delta _ {jj'} /2]$.
Substituting the expansions of electric and magnetic fields (\ref{eq:E},\ref{eq:H}),
we can write
\begin{align}
  \expval{F} = 
  \begin{pmatrix}
    \mathbb{E} _ {\vb{k}} ^ {(3)+}
    \\
    \mathbb{E} _ {\vb{k}} ^ {(3)-}
  \end{pmatrix} ^ \dagger
  \begin{pmatrix}
    \mathsf{F} _ {\vb{k}} ^ {++} & \mathsf{F} _ {\vb{k}} ^ {+-} 
    \\
    \mathsf{F} _ {\vb{k}} ^ {-+} & \mathsf{F} _ {\vb{k}} ^ {--} 
  \end{pmatrix}
  \begin{pmatrix}
    \mathbb{E} _ {\vb{k}} ^ {(3)+}
    \\
    \mathbb{E} _ {\vb{k}} ^ {(3)-}
  \end{pmatrix}
  \label{eq:F}
\end{align}
where the matrix element reads
\begin{align}
  \qty[\mathsf{F} _ {\vb{k}} ^ {\sigma \sigma'}] _ {n n'} 
  &= \frac{1}{\Omega/g} \frac{1}{Z _ 0} 
  \frac{(\sigma K _ {\vb{k} _ {n}} ^ {(3)} + \sigma' K _ {\vb{k} _ {n}} ^ {(3)})/2}{n\Omega/c} 
  \delta _ {n n'} 
  \label{eq:F _ nn'}
\end{align}
From the expression \eqref{eq:F _ nn'}, it is clear that there is no lateral force ($F = 0$) if there is no spatial modulation ($g = 0$).
In other words, the radiation emitted by the modulated surface does not possess any lateral momentum and cannot drag the second surface.
It is also evident from Eq.~\eqref{eq:F _ nn'} that the off-diagonal contributions ($\sigma \neq \sigma'$) stem from evanescent waves while the diagonal ones ($\sigma = \sigma'$) from propagating waves as in the power loss.
We can associate the lateral force with the power loss as we can write
\begin{align} 
  \qty[\mathsf{P} _ {\vb{k} (3)} ^ {\sigma \sigma'}] _ {n n'}
  &= \frac{v _ \mathrm{ph}}{Z _ 0/Z _ {\vb{k} _ n} ^ {(3)}}
  \qty[\mathsf{F} _ {\vb{k}} ^ {\sigma \sigma'}] _ {n n'}.
  \label{eq:F_vs_P}
\end{align}
Remind that $v _ \mathrm{ph} = \Omega/g$ is the peristaltic velocity.
This relation is reminiscent of the power loss as a product of velocity and a frictional force.
In particular, for the propagating waves ($\operatorname{Im} K _ {\vb{k}} ^ {(3)} = 0$),
the wave impedance recovers the one in free space, $Z _ 0 / Z _ \mathbf{k} ^{(3)} = 1$,
and 
$\mathsf{P} _ {\vb{k} (3)} ^ {\sigma \sigma'}=
v _ \mathrm{ph} \mathsf{F} _ {\vb{k}} ^ {\sigma \sigma'}$.
As the grating dressed by the electromagnetic radiation requires the power given by $\mathsf{P} _ {\vb{k}(\tau)} ^ {\sigma \sigma'}$ to travel at the peristaltic velocity $v _ \mathrm{ph}$,
the relation can indeed be viewed as the power loss as the product of the peristaltic velocity and noncontact friction between the grating and the second surface.

\begin{figure}[tbp]
  \centering
  \includegraphics[width=\linewidth]{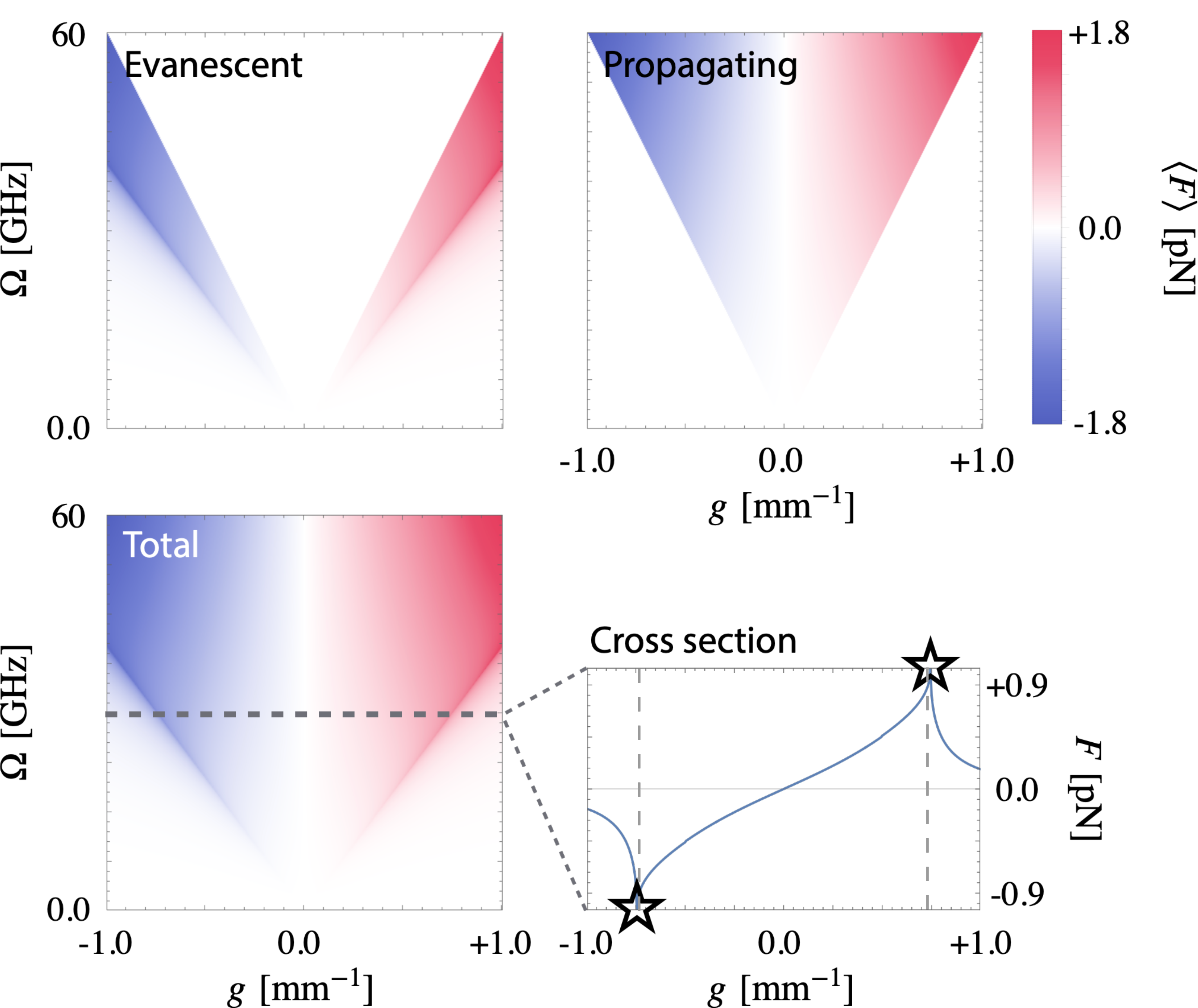}
  \caption{
    Frictional force $F$ per unit area on the second surface.
    The contributions from evanescent waves (top left), the ones from propagating waves (top right), and the total contribution (bottom left) are shown.
    The bottom right figure is a cross-section at the grey dashed line in the bottom left plot.
    The directions of the forces coincide with the signature of the modulation wavenumber $g$.
    The total force is peaked at the luminal condition in the dielectric side ($|\Omega/g| = c/\sqrt{\epsilon _ 3}$),
    where \v{C}erenkov radiation emerges in the vacuum region as in the case of corrugated surfaces \cite{oue2022cerenkov}.
    We have used the same parameters as in the previous figures to produce these figures (\figref{fig:loss}).
  }
  \label{fig:force}
\end{figure}
As we did in the power loss calculation, we can substitute the field amplitudes evaluated in the previous steps (\ref{eq:E_multiplied}--\ref{eq:E_multiplied_outer3}) into the quadratic expression \eqref{eq:F} to compute the force of interest.
In \figref{fig:force}, we show the lateral force as a function of the grating parameters, $g$ and $\Omega$.
First, we can confirm the force vanishes at $g=0$,
where the grating provides no momentum.
Second, the direction of the force corresponds to the signature of $g$,
and it seems that the second surface feels and is dragged by the radiation emitted from the grating.
In other words, the modulation induces an electromagnetic force in the direction of the modulation velocity $v _ \mathrm{ph}$.
Third, the force is peaked at the luminal condition $v _ \mathrm{ph} = c/\sqrt{\epsilon _ 3}$ in the dielectric substrate (star symbols in the cross section plot).
One reason is that the radiation strongly interacts with the grating and is amplified at the condition, as we have also discussed in the power loss.
Another reason is that the emission is in the horizontal direction at the luminal condition;
thus, the radiation has a large momentum parallel to the surfaces.

\section{Discussion}
To sum up, in this work, we have analysed the electromagnetic properties of the space-time-modulated grating, which performs peristaltic motion.
Under a DC voltage, the grating emits electromagnetic radiation as it provides not only momentum $g$ but also energy $\Omega$ to the electromagnetic field, and the DC input voltage is converted to AC fields.
The field emission from the grating induces noncontact frictional forces.
The noncontact friction can be viewed as a consequence of the energy loss as in the conventional friction.
In our case, this can be seen in the force-loss relation \eqref{eq:F_vs_P}, where the product of the friction and the peristaltic speed gives the energy loss.

In the present setup, we can tune the effective surface permittivity $\epsilon _ 1$, the modulation parameters (strength $\alpha$ and frequencies $\mathbf{q}$) and the gap width between the two surfaces.
As described around \figref{fig:force}, the force is efficiently exerted at the luminal condition ($|g| = \Omega/c$) because the emitted radiation travels parallel to the grating so that it can strongly interact with the grating to gain much lateral momentum, which is delivered to the top layer.
This is one of the crucial conditions to maximise the force.
With that condition satisfied, we can increase the emitted radiation intensity to enlarge the force.
The emitted radiation intensity increases as the time variation of the surface permittivity becomes significant. 
Remind that the radiation flux is equivalent to the power loss, $P = \vec{\mathcal{E}} \cdot \epsilon _ 0 \dot\epsilon \vec{\mathcal{E}} \propto \epsilon _ 1 \alpha \Omega$.
Thus, the force can be enlarged by adopting materials with large permittivity $\epsilon _ 1$ and making the modulation amplitude $\alpha$ and temporal frequency $\Omega$ as large as possible.

\begin{figure}[htbp]
  \centering
  \includegraphics[width=.7\linewidth]{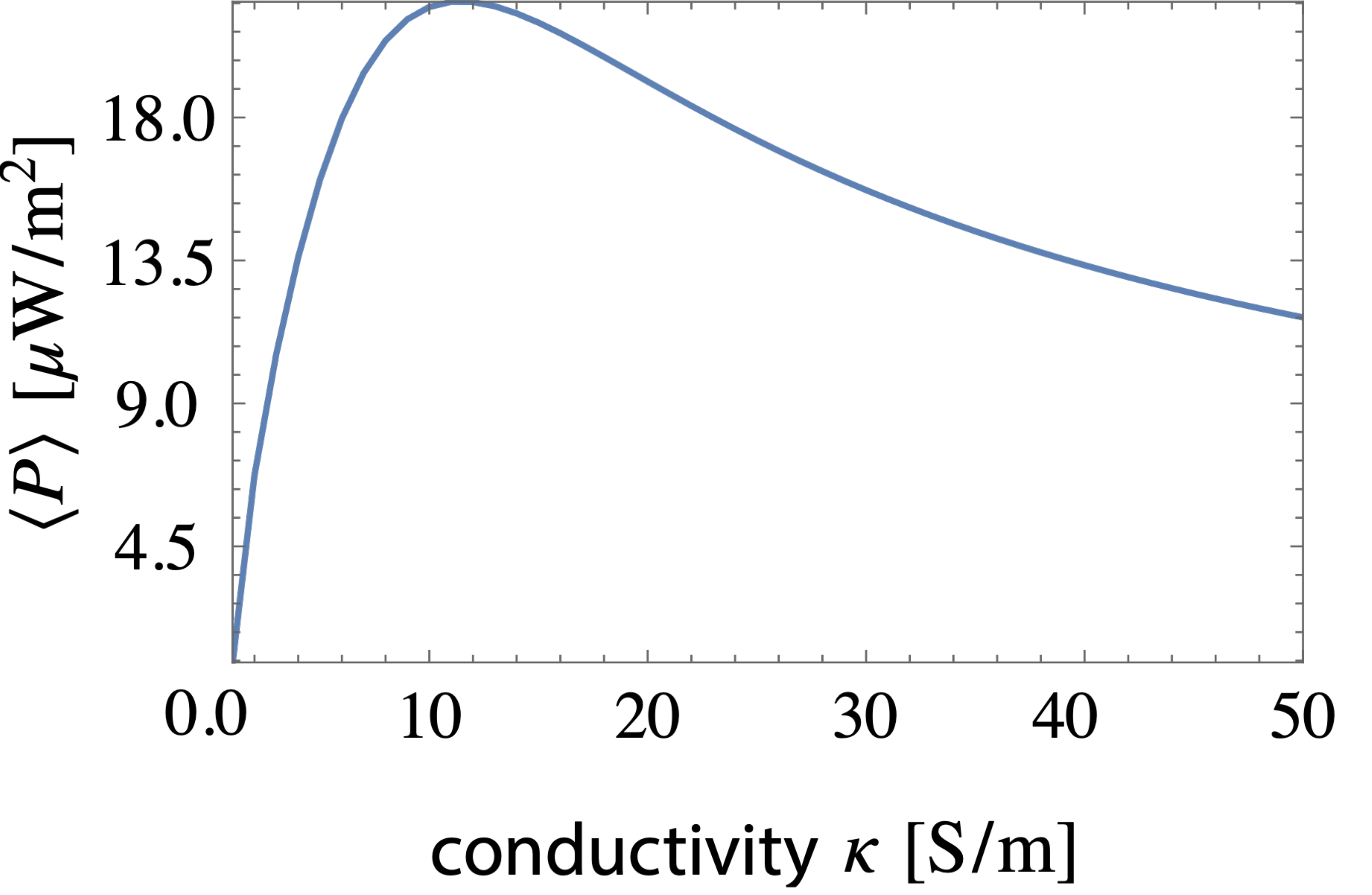}
  \caption{
    Power loss as a function of the conductivity of the upper medium.
    The loss goes small if the conductivity is small or very large.
    This is because nothing is absorbed if the medium is far less conducting or perfectly conducting.
    If the medium is much less conducting (perfectly conducting), every radiation can be transmitted through (reflected by) the medium.
    We used the following parameters to generate this plot:
    $g = 2.0\ \mathrm{mm ^ {-1}}$;
    $\alpha = -0.19$; 
    $\epsilon_1 = 3.3 \times 10^{-6}$;
    $E _ \mathrm{in} = 0.03\ \mathrm{V/mm}$;
    $\Omega = 0.8 gc/\sqrt{\epsilon _ 3}$.
  }
  \label{fig:p-kappa}
\end{figure}

Another parameter that we can adjust is the conductivity $\kappa$ of the top material.
If the medium is much less conductive, little radiation is absorbed there; hence, the amount of momentum transfer (i.e.~the frictional force) will be much weaker.
At the perfectly resistive limit ($\kappa \rightarrow 0$), no radiation will be absorbed, and there will be no force.
In the opposite limit, where the medium is perfectly conducting ($\kappa \rightarrow \infty$), any radiation cannot enter the medium and is reflected so that there will be no force.
From these discussion, we can expect that there is an optimal conductivity where the dissipation is not overkill, and the force is maximised.
In order to find such an optimal condition to dissipate the emitted radiation, we have calculated the power loss $\expval{P}$ as a function of the conductivity $\kappa$ (see \figref{fig:p-kappa}).
The power loss goes small as the conductivity becomes very small or very large as expected.
The optimal conductivity in the present case is $\kappa \approx 11\ \mathrm{[S/m]}$, which is the number we have employed in FIGs.~\ref{fig:loss} and \ref{fig:force}.

\begin{figure}[htbp]
  \centering
  \includegraphics[width=.8\linewidth]{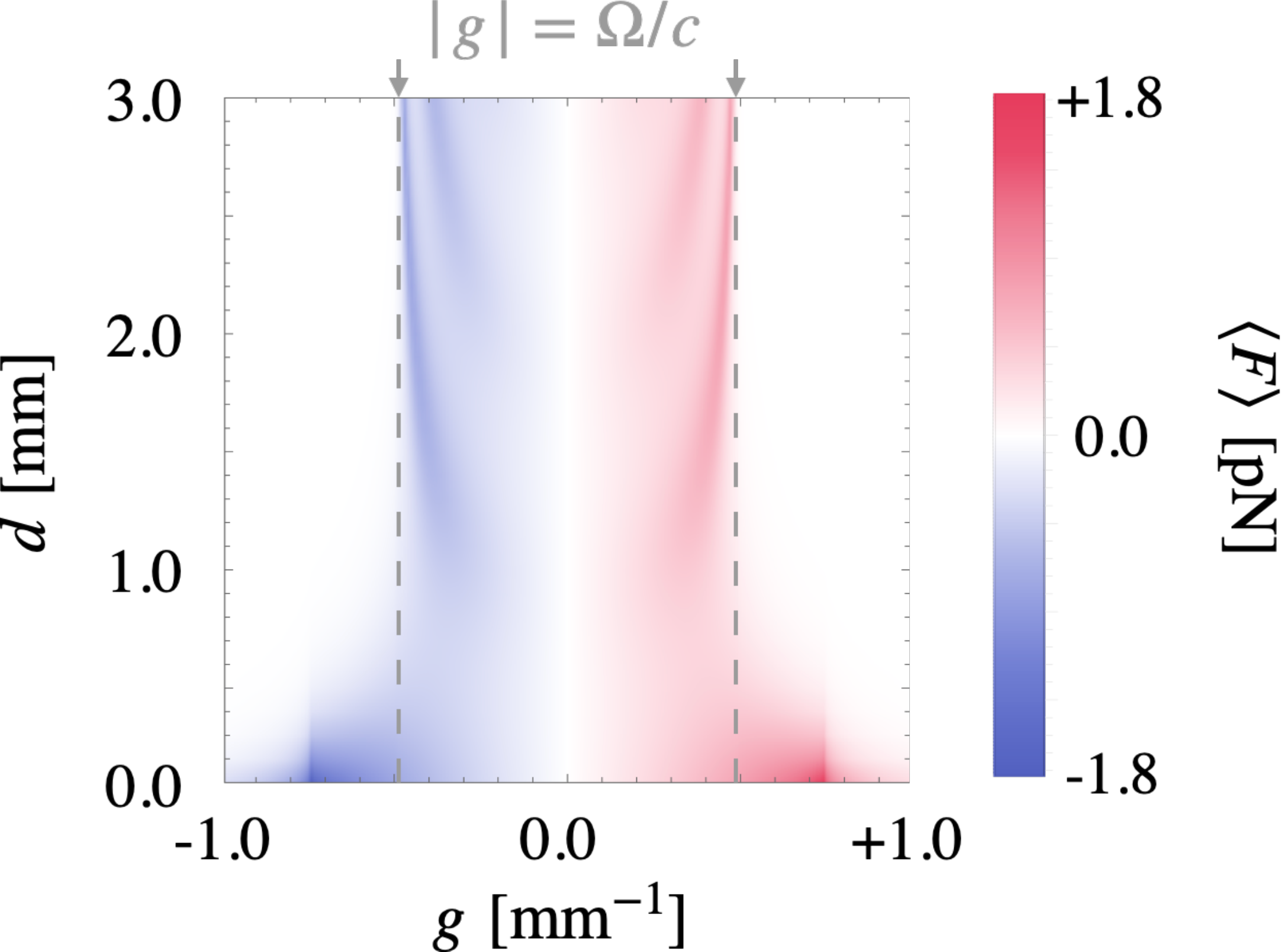}
  \caption{
    Frictional force as a function of the spatial modulation frequency $g$ and the gap width $d$ for a given modulation frequency $\Omega = 30\ \mathrm{GHz}$.
    The other parameters are the same as the ones used in the main text:
    $\kappa = 11\ \mathrm{S/m}$;
    $\alpha = -0.19$; 
    $\epsilon_1 = 3.3 \times 10^{-6}$;
    $E _ \mathrm{in} = 0.03\ \mathrm{V/mm}$.
    The vertical dashed lines represents the luminal condition $|g| = \Omega/c$ for the gap region.
  }
  \label{fig:f-g-d}
\end{figure}

Additionally, in \figref{fig:f-g-d}, we show the frictional force as a function of the spatial modulation frequency $g$ and the gap width $d$ between the two surfaces for a given temporal modulation frequency.
We can recognise the force in the subluminal regime ($|g| > \Omega/c $) decays as the gap width increases.
This is because the electromagnetic radiation is evanescent one in this region as shown in FIG.~3 in the main text.
On the other hand, in the superluminal regime ($|g| < \Omega/c$), the emitted radiation is propagating, and the force does not decay unlike in the subluminal case.
This is why we can recognise that fringe appears as the gap width becomes large. 
The fringe can be attributed to radiation confined between the surfaces (Fabry-P\'{e}rot-type modes).

Our system is similar to the electrostatic friction~\cite{persson1998theory,stipe2001noncontact, chumak2004effects, kuehn2006dielectric, den2015spin, de2017dissipation, heritier2021spatial} in the ensuing aspect.
The electrostatic friction is mediated by a static field arises from electric charges or polarisations.
Similarly, in our scenario, the static electric polarisation induced by the DC voltage brings about a electrostatic field, resulting in friction.
On the other hand, our configuration is different from the electrostatic case in that our system engages in peristaltic, rather than physical, motion.
As this motion is non-physical, the resultant velocity can surpass that of light, and the mediator field can be of the \v{C}erenkov variety, which is no longer static but dynamical.
Consequently, altering the distance between the two objects produces an oscillatory effect.

As we studied previously~\cite{oue2021calculating,oue2022cerenkov},
the scattering from an infinitesimally thin grating is much the same as that from a shallow groove.
In this sense, the noncontact frictional force due to the peristaltic motion,
which is investigated in the present work,
can be experimentally tested by acoustically deforming a dielectric surface~\cite{chenu1994giant,issenmann2006bistability, issenmann2008deformation, rambach2016visualization} as well as 
utilising a field-programmable gate array~\cite{zhang2018space,zhang2019breaking,li2021programmable}.
Modulating the conductivity of graphene on a substrate \cite{chen2011controlling,li2014ultrafast,liu2016time,poumirol2017electrically,tasolamprou2019experimental} is another possible route for the experimental implementation as it can be effectively regarded as an infinitesimally thin sheet \cite{gonccalves2016introduction}.
The force detection can be done by putting a metallic probe in close proximity of the surface as in the case of electrostatic friction~\cite{stipe2001noncontact, kuehn2006dielectric, den2015spin}.

\begin{acknowledgments}
  The authors thank M.~G.~Silveirinha for fruitful discussions.
  D.O.~is supported by JSPS Overseas Research Fellowship, by the Institution of Engineering and Technology (IET), and by Funda\c{c}\~ao para a Ci\^encia e a Tecnologia and Instituto de Telecomunica\c{c}\~oes under project UIDB/50008/2020.
  J.B.P.~acknowledges support from the Gordon and Betty Moore Foundation.
  K.D.~acknowledges support from Natural Science Foundation of China (No.~12174072) and Natural Science Foundation of Shanghai (No.~21ZR1403700).
\end{acknowledgments}

\appendix
\section{Reflection and transmission matrices}
Here, we describe how to obtain the reflection and transmission matrices of the space-time-modulated grating.
First, as mentioned in the main text,
we need to match the boundary condition at $z = 0$ in the presence of the grating-induced source.

The tangential component of the electric field is continuous at the grating plane as usual,
\begin{align}
  \lim _ {\delta h \downarrow 0} \vec{u} _ y \cdot [\vec{\mathcal{E}}(\mathbf{x}, +\delta h) - \vec{\mathcal{E}}(\mathbf{x},-\delta h)] = 0.
  \tag{\ref{eq:E_continuity}}
\end{align}
On the other hand,
the magnetic field is discontinuous because the infinitesimally thin grating provides a surface permittivity,
\begin{align}
  \lim _ {\delta h \downarrow 0} \vec{u} _ x \cdot [\vec{\mathcal{H}}(\mathbf{x}, +\delta h) - \vec{\mathcal{H}}(\mathbf{x},-\delta h)] =
  \vec{u} _ y \cdot \frac{\dot{\epsilon} _ \mathrm{sf} \vec{\mathcal{E}} (\mathbf{x},0)}{cZ _ 0}.
  \tag{\ref{eq:H_continuity}}
\end{align}

\begin{widetext}
Applying the Fourier transform to these continuity equations (\ref{eq:E_continuity}, \ref{eq:H_continuity}) in the real space and the time domain,
we have corresponding simultaneous equations in the reciprocal space and the frequency domain,
\begin{align}
    \qty(
    \mathbb{E} _ {\vb{k}} ^ {(3)+}
    +\mathbb{E} _ {\vb{k}} ^ {(3)-}
    )
    -\qty(
    \mathbb{E} _ {\vb{k}} ^ {(2)+}
    +\mathbb{E} _ {\vb{k}} ^ {(2)-}
    ) = 0,
    \label{eq:E_continuity_FT}
\end{align}
\begin{align}
    \qty(
    \mathsf{N} _ {\vb{k}} ^ {(3)+} \mathbb{E} _ {\vb{k}} ^ {(3)+}
    +\mathsf{N} _ {\vb{k}} ^ {(3)-} \mathbb{E} _ {\vb{k}} ^ {(3)-}
    )
    -\qty(
    \mathsf{N} _ {\vb{k}} ^ {(2)+} \mathbb{E} _ {\vb{k}} ^ {(2)+}
    +\mathsf{N} _ {\vb{k}} ^ {(2)-} \mathbb{E} _ {\vb{k}} ^ {(2)-}
    ) = \mathsf{L} _ {\vb{k}} 
    \qty(
    \mathbb{E} _ {\vb{k}} ^ {(2)+}
    +\mathbb{E} _ {\vb{k}} ^ {(2)-}
    ),
    \label{eq:H_continuity_FT}
\end{align}
where the $\mathsf{N}$ matrices are diagonal while the $\mathsf{L}$ matrix is responsible for the grating contribution and possesses off-diagonal elements that trigger the frequency shifts,
\begin{align}
  \qty[\mathsf{N}_{\vb{k}}^{(\tau)\sigma}] _ {nn'}
  = 
  \frac{\sigma K_{\vb{k}_ {n'}}^{(\tau)}}{n'\Omega/c}
  \delta_{nn'},
  \quad
  \qty[\mathsf{L}_{\vb{k}}] _ {nn'}
  = \frac{\Omega}{c} \epsilon _ 1
  \qty{
    \alpha 
    (n'-1) \delta_{n,n'-1}
    +
    \alpha 
    (n'+1) \delta_{n,n'+1}
    + n' \delta_{nn'}
  }.
\end{align}

Rearranging Eqs.~(\ref{eq:E_continuity_FT}, \ref{eq:H_continuity_FT}) into a matrix form,
we can obtain
\begin{align}
  \begin{pmatrix}
    +1 & -1\\
    +\mathsf{N} _ {\vb{k}} ^ {(3)+} & -(\mathsf{N} _ {\vb{k}}^ {(2)-} + \mathsf{L} _ {\vb{k}})
  \end{pmatrix}
  \begin{pmatrix}
    \mathbb{E} _ {\vb{k}} ^ {(3)+}\\
    \mathbb{E} _ {\vb{k}} ^ {(2)-}
  \end{pmatrix}
  =  
  \begin{pmatrix}
    -1 & +1\\
    -\mathsf{N} _ {\vb{k}} ^ {(3)-} & +(\mathsf{N} _ {\vb{k}}^ {(2)+} + \mathsf{L} _ {\vb{k}})
  \end{pmatrix}
  \begin{pmatrix}
    \mathbb{E} _ {\vb{k}} ^ {(3)-}\\
    \mathbb{E} _ {\vb{k}} ^ {(2)+}
  \end{pmatrix}.
  \label{eq:sca_eq}
\end{align}
Inverting the matrix on the left-hand side in Eq.~\eqref{eq:sca_eq},
we can write
\begin{align}
  \begin{pmatrix}
    \mathbb{E} _ {\vb{k}} ^ {(3)+}\\
    \mathbb{E} _ {\vb{k}} ^ {(2)-}
  \end{pmatrix}
  &=  
  \begin{pmatrix}
    +1 & -1\\
    +\mathsf{N} _ {\vb{k}} ^ {(3)+} & -(\mathsf{N} _ {\vb{k}}^ {(2)-} + \mathsf{L} _ {\vb{k}})
  \end{pmatrix} ^ {-1}
  \begin{pmatrix}
    -1 & +1\\
    -\mathsf{N} _ {\vb{k}} ^ {(3)-} & +(\mathsf{N} _ {\vb{k}}^ {(2)+} + \mathsf{L} _ {\vb{k}})
  \end{pmatrix}
  \begin{pmatrix}
    \mathbb{E} _ {\vb{k}} ^ {(3)-}\\
    \mathbb{E} _ {\vb{k}} ^ {(2)+}
  \end{pmatrix},
  \\
  &:= 
  \begin{pmatrix}
    \mathsf{R} _ {\vb{k}} & \mathsf{T} _ {\vb{k}} \\
    \mathsf{T} _ {\vb{k}}' & \mathsf{R} _ {\vb{k}}' 
  \end{pmatrix}
  \begin{pmatrix}
  \mathbb{E} _ {\vb{k}} ^ {(3)-}\\
  \mathbb{E} _ {\vb{k}} ^ {(2)+}
  \end{pmatrix}.
  \tag{\ref{eq:scattering_matrix}}
\end{align}
\end{widetext}

\section{Multiple reflection}
\begin{figure}[htbp]
  \centering
  \includegraphics[width=.6\linewidth]{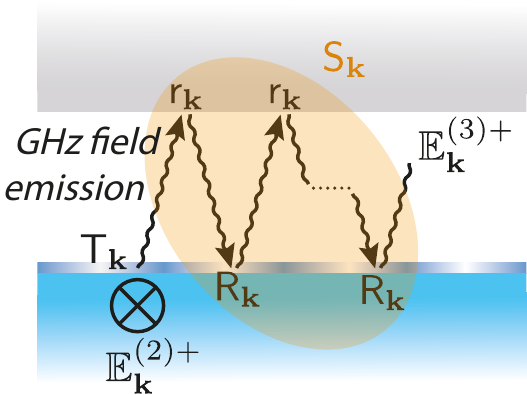}
  \caption{Multiple reflections between two surfaces.
    The DC electrostatic input (normal to the plane of the figure) is multiplied by the transmission matrix of the grating, 
    $\mathbb{E} _ \mathbf{k} ^ \mathrm{sou} = \mathsf{T} _ \mathbf{k} \mathbb{E} _ \mathbf{k} ^ {(2)+}$,
    which is fed into the gap region as a source of multiple reflections between the upper flat surface and the grating.
    The resultant field in the gap region is the product of the source field $\mathbb{E} _ \mathbf{k} ^ \mathrm{sou}$ and the multiple reflection factor,
    $\mathsf{S} _ \mathbf{k}  = (1-\mathsf{R} _ \mathbf{k} \mathsf{r} _ \mathbf{k} )^{-1}$,
    where $\mathsf{R} _ \mathbf{k} $ and $\mathsf{r} _ \mathbf{k} $ are the reflection matrices of the lower and upper surfaces.
  }
  \label{fig:multiple}
\end{figure}

Making use of the multiple reflection arguments depicted in \figref{fig:multiple},
we can expand the field amplitude in powers of the reflection matrices.
The upgoing field between the surfaces is composed of fields reflected even times,
\begin{align}
  \mathbb{E} _ {\vb{k}} ^ {(3)+} =
  (1 + \mathsf{R} _ {\vb{k}} \mathsf{r} _ {\vb{k}} +
  \mathsf{R} _ {\vb{k}} \mathsf{r} _ {\vb{k}} \mathsf{R} _ {\vb{k}} \mathsf{r} _ {\vb{k}} + \cdots)
  \mathsf{T} _ {\vb{k}} \mathbb{E} _ {\vb{k}} ^ {(2)+},
\end{align}
while the downgoing field is reflected odd times,
\begin{align}
  \mathbb{E} _ {\vb{k}} ^ {(3)-} =
  (\mathsf{r} _ {\vb{k}}  + \mathsf{r} _ {\vb{k}} \mathsf{R} _ {\vb{k}} \mathsf{r} _ {\vb{k}} +
  \mathsf{r} _ {\vb{k}} \mathsf{R} _ {\vb{k}} \mathsf{r} _ {\vb{k}} \mathsf{R} _ {\vb{k}} \mathsf{r} _ {\vb{k}} + \cdots)
  \mathsf{T} _ {\vb{k}} \mathbb{E} _ {\vb{k}} ^ {(2)+},
\end{align}
where $\mathsf{r} _ {\vb{k}}$ is the reflection matrix of the upper flat surface whose elements are nothing but the Fresnel coefficients,
\begin{align}
  [\mathsf{r} _ {\vb{k}}] _ {nn'} =
  \frac{K _ {\vb{k} _ n}  ^ {(3)} - K _ {\vb{k} _ n} ^ {(4)}}{K _ {\vb{k} _ n}  ^ {(3)} + K _ {\vb{k} _ n} ^ {(4)}}
  e ^ {2i K _ {\vb{k} _ n}  ^ {(3)} d}
  \delta _ {nn'}.
\end{align}
Note that we should put the exponential factor $e ^ {2i K _ {\vb{k} _ n}  ^ {(3)} d}$ to incorporate the field propagation and the decay in the gap region.
Defining the multiple reflection factor,
$\mathsf{S} _ {\vb{k}} := (1-\mathsf{R} _ {\vb{k}} \mathsf{r} _ {\vb{k}})^{-1}$,
we can write compactly as
\begin{align}
  \mathbb{E} _ {\vb{k}} ^ {(3)+} =
  \mathsf{S} _ {\vb{k}} \mathsf{T} _ {\vb{k}} \mathbb{E} _ {\vb{k}} ^ {(2)+},
  \quad
  \mathbb{E} _ {\vb{k}} ^ {(3)-} =
  \mathsf{r} _ {\vb{k}} \mathsf{S} _ {\vb{k}} \mathsf{T} _ {\vb{k}} \mathbb{E} _ {\vb{k}} ^ {(2)+}.
  \tag{\ref{eq:E_multiplied}}
\end{align}
The corresponding fields in the upper and lower semi-infinite regions are
\begin{align}
  \mathbb{E} _ {\vb{k}} ^ {(4)+} &=
  \mathsf{t} _ {\vb{k}} \mathsf{S} _ {\vb{k}} \mathsf{T} _ {\vb{k}} 
  \mathbb{E} _ {\vb{k}} ^ {(2)+},
  \tag{\ref{eq:E_multiplied_outer4}}
  \\
  \mathbb{E} _ {\vb{k}} ^ {(2)-} &=
  (\mathsf{R} _ {\vb{k}} ' + \mathsf{T} _ {\vb{k}}' \mathsf{r} _ {\vb{k}} \mathsf{S} _ {\vb{k}} \mathsf{T} _ {\vb{k}})
  \mathbb{E} _ {\vb{k}} ^ {(2)+},
  \tag{\ref{eq:E_multiplied_outer3}}
\end{align}
where the transmission matrix $\mathsf{t} _ \mathbf{k}$ of the upper flat surface is composed of the Fresnel coefficients,
\begin{align}
  \qty[\mathsf{t} _ \mathbf{k}] _ {nn'} =
  \frac{2 K _ {\vb{k} _ n}  ^ {(3)}}{K _ {\vb{k} _ n}  ^ {(3)} + K _ {\vb{k} _ n} ^ {(4)}}
  e ^ {i (K _ {\vb{k} _ n}  ^ {(3)} - K _ {\vb{k} _ n}  ^ {(4)})d}
  \delta _ {nn'}.
\end{align}

\bibliography{%
  bib/all,%
  textbook/textbook_all%
}
\end{document}